\documentclass[conference]{IEEEtran}
\IEEEoverridecommandlockouts
\usepackage{amsmath,amssymb,amsfonts}
\usepackage{algpseudocode}
\usepackage{algorithm}
\usepackage{graphicx}
\usepackage{svg}
\usepackage{booktabs}
\usepackage{quantikz}
\usepackage{siunitx}

\usepackage[hidelinks]{hyperref}
\hypersetup{
     colorlinks=true,
     linkcolor=blue,
     citecolor=black,
     urlcolor=cyan,
}

\usepackage[numbers,sort&compress]{natbib}
\usepackage{braket}
\usepackage{textcomp}
\usepackage{xcolor}
\def\BibTeX{{\rm B\kern-.05em{\sc i\kern-.025em b}\kern-.08em
    T\kern-.1667em\lower.7ex\hbox{E}\kern-.125emX}}
    
\begin{document}

\title{A Quantum Range-Doppler Algorithm for Synthetic Aperture Radar Image Formation\\}

\author{\IEEEauthorblockN{Alessandro Giovagnoli}
\IEEEauthorblockA{\textit{German Aerospace Center (DLR)} \\
Oberpfaffenhofen, Germany\\
alessandro.giovagnoli@dlr.de
}
\and
\IEEEauthorblockN{Sigurd Huber}
\IEEEauthorblockA{\textit{German Aerospace Center (DLR)} \\
Oberpfaffenhofen, Germany\\
sigurd.huber@dlr.de
}
\and
\IEEEauthorblockN{Gerhard Krieger}
\IEEEauthorblockA{\textit{German Aerospace Center (DLR)} \\
Oberpfaffenhofen, Germany\\
gerhard.krieger@dlr.de
}
}

\maketitle

\begin{abstract}
	Synthetic aperture radar (SAR) is a well established technology in the field of Earth remote sensing. Over the years, the resolution of SAR images has been steadily improving and the pixel count increasing as a result of advances in the sensor technology, and so have the computational resources required to process the raw data to a focused image. Because they are a necessary step in the study of the retrieved data, new high-resolution and low-complexity focusing algorithms are constantly explored in the SAR literature. The theory of quantum computing proposes a new computational framework that might allow to process a vast amount of data in a more efficient way. Relevant to our case is the advantage proven for the quantum Fourier transform (QFT), the quantum counterpart of a fundamental element of many SAR focusing algorithms. Motivated by this, in this work we propose a quantum version of the range-Doppler algorithm. We show how in general reference functions, also referred to as processing kernels or focusing filters, and a key element in many SAR focusing algorithms, can be mapped to quantum gates. We present the quantum circuit performing the SAR raw data focusing and we discuss in detail its computational complexity. We find that the core of the proposed quantum range-Doppler algorithm has a computational complexity, namely the number of single- and two-qubit gates, of $O(N)$, which is less than its classical counterpart with computational complexity $O(N \log N)$.
\end{abstract}

\section{Introduction}
Synthetic Aperture Radar (SAR) is a technique which allows imaging the Earth's surface independent of daylight and weather conditions at microwave frequencies. The retrieved data, which consist of back-scattered radio waves, can be collected and processed over a synthetic aperture, allowing for a higher resolution beyond the physical size of the antenna. Exploiting the phase history and Doppler shifts, a two- or three-dimensional image of the terrain or target can be reconstructed \cite{6504845}. The raw data, which can be interpreted as an image, consist of an array of complex values encoding the magnitude and phase of the collected waves at each time and space step. 

Depending on the acquisition mode of the SAR system, different algorithms can be employed to process the initial raw data and obtain the focused image \cite{cumming2005digital}. According to publicly available datasets of recent SAR missions \cite{1435823, TanDEM, ALOS, NISAR, 9944234, SENTINEL}, the size of SAR images, which, for simplicity, we assume to be a square complex-valued matrix of $N$ total pixels, can vary between $N = 5000^2$ and $N = 50000^2$, depending on the acquisition mode.

Since high-resolution and more precise focusing algorithms require a higher computational overhead \cite{rs14051258}, one has to choose the desired trade-off between quality and computational resources. For this reason, a major area of research in the SAR literature focuses on increasingly efficient processing algorithms. New techniques have recently been proposed, such as autofocus to reduce phase errors \cite{6471225}, artificial intelligence models to focus the raw data \cite{9173392} or compressive sensing techniques \cite{5414307}, which leverage signal sparsity to obtain high-resolution images with a lower number of measurements.

Justified by the potential of a lower computational cost, researchers have started to look into the quantum information theory paradigm to tackle signal processing tasks \cite{NorouziLarki2023, arxiv1, arxiv2} driven by the promising results of the quantum Fourier transform, which is proven to have a lower computational complexity compared to its classical counterpart \cite{Nielsen_Chuang_2010}. In the present work we address the task of lowering the computational complexity of SAR raw data processing exploiting the quantum computing framework. In particular, we will focus on the range-Doppler algorithm \cite{cumming2005digital}, which is historically the first SAR processing algorithm proposed, although it is today still widely used. A quantum version of the omega-k algorithm has been presented in \cite{Waller2023}. In this work we propose a quantum version of the range-Doppler algorithm with the goal of minimizing the number of qubits and the gate count. In particular, emphasis is placed on the complexity of the gate decomposition and its classical preprocessing. Parallel to our work, also in \cite{2504.01832} a quantum adaptation of the range-Doppler algorithm has been explored, however without a thorough analysis of relative quantum and classical complexity.

The rest of this paper is organized as follows: in section \ref{sec:classical-alg} we briefly present the classical range-Doppler algorithm and highlight the approximation that allows mapping the problem into the quantum computing framework. In section \ref{sec:quantum-alg} we show in detail the proposed quantum range-Doppler algorithm. In section \ref{sec:results} we show the results obtained by applying the proposed quantum algorithm to a simulated example. In section \ref{sec:complexity} we study in detail the computational complexity of the proposed algorithm, highlighting, where necessary, also the classical preprocessing.

\section{Classical Range-Doppler Algorithm}\label{sec:classical-alg}
The range-Doppler algorithm is a well-known processing technique in SAR to obtain a focused image from the raw signals. It consists of three main steps: compressing the raw signal along the range dimension, correcting for the range cell migration (RCM), and compressing along the azimuth dimension.  
The compressions are performed by convolving the received signals with the appropriate reference functions, which are the complex conjugate of the functions that describe the signal that we expect to receive. By the convolution theorem, this operation can also be carried out in the frequency domain, by multiplying the Fourier transform of the signal by the Fourier transform of the reference function. 
The RCM correction is, instead, usually performed through a range interpolation. Here, we will assume that the RCM is range-invariant, which is true if the range region is not too wide \cite{cumming2005digital}. Under this assumption, one can also correct for the RCM with an appropriate reference function in the frequency domain.

As is usually done, we assume the transmitted waveform to be a chirp, which exhibits a linearly increasing frequency in time with a constant rate $\alpha$ and a pulse length $T$. After scattering on ground the received signal for a monostatic setup is

\begin{equation}\label{eq:signal-raw-data}
	\begin{aligned}
		s(\tau, \eta)
		=
		&A_0 a(\eta)
		\chi \left(\tau - \frac{2r(\eta)}{c}\right) 
		\exp
		\left\{\mathrm{i} \pi \alpha \left(\tau - \frac{2r(\eta)}{c}\right)^2 
		\right\} 
		\\
		& \exp
		\left\{ -2 \pi \mathrm{i} f_0 \frac{2r(\eta)}{c} 
		\right\}
		,
	\end{aligned}
\end{equation}
where $\tau$ and $\eta$ are, respectively, the fast and slow time, which are the times along the range and  azimuth directions. $A_0$ accounts for quasi constant factors such as the magnitude of the transmitted waveform, propagation losses, etc. $a(\eta)$ is the antenna pattern, $\chi$ is a box function setting the length of the chirp to $T$ since $\chi(\eta) = 1$ if $\eta \in [-T/2, T/2]$ and $0$ otherwise, $r(\eta)$ is the range distance between the sensor and a target on ground, $f_0$ is the center frequency of the transmitted waveform and $c$ the speed of light. For simplicity, we set in the following $a(\eta) = 1$.

The reference functions for range compression, RCM correction and azimuth compression are, respectively,

\begin{equation}\label{eq:reference-functions}
	\begin{aligned}
		h_\mathrm{r}(\tau) 
		&=
		\chi(\tau)
		\exp\left\{- \mathrm{i} \pi \alpha \tau^2\right\}
		\\
		h_\mathrm{m}(f_\tau, \eta)
		&=
		\exp\left\{
		2 \pi \mathrm{i} f_\tau \frac{2 \Delta r(\eta)}{c} 
		\right\}
		\\
		h_\mathrm{a}(\tau, \eta)
		&=
		\exp\left\{
		2 \pi \mathrm{i} f_0 \frac{2r(\eta)}{c} 
		\right\}
		,
		\\
	\end{aligned}
\end{equation}
where, if we indicate with $r_0$ the slant range, which is the line-of-sight distance from the radar to the target, we can approximate $r(\eta)$ as

\begin{equation}\label{eq:taylor-approx}
	r(\eta) = \sqrt{r_0^2 + v^2 \eta^2} \approx r_0 + \frac{v^2 \eta^2}{2 r_0} = r_0 + \Delta r(\eta)
	,
\end{equation}
where $v$ is the speed of the SAR sensor.
Their Fourier transforms $H_\mathrm{r}(f_\tau)$, $H_\mathrm{m}(f_\tau, f_\eta)$ and $H_\mathrm{a}(\tau, f_\eta)$ have to be computed to carry out the convolutions in frequency domain. Rather than explicitly carrying out the Fourier transform of these functions, we will use the principle of stationary phase (POSP), which is a well-known approximation in the theory of oscillatory integrals \cite{POSP}. The POSP allows evaluating the integral of a rapidly oscillating function by approximating its value at the critical points of the phase of the integrand. 
More precisely, given a function of the form $e^{i \theta(x)}$, we can approximate its integral as

\begin{equation}\label{eq:POSP}
	\int_{\mathbb{R}} e^{\mathrm{i} \theta(x)} dx \approx\sqrt{\frac{2 \pi}{|\theta''(x_0)|}} e^{\text{sign}(\theta''(x_0))\mathrm{i} \frac{\pi}{4}} \cdot e^{\mathrm{i}\theta(x_0)} \approx e^{\mathrm{i} \theta(x_0)}
	,
\end{equation}
with $x_0$ such that $\theta'(x_0) = 0$. The first approximation in equation (\ref{eq:POSP}) is due to the POSP, and the second to the fact that the neglected terms represent global factors, and for this reason we can ignore them \cite{cumming2005digital}. The reason why we are interested in using the POSP is that it guarantees that $H_\mathrm{r}(f_\tau), H_\mathrm{m}(f_\tau, f_\eta)$ and $H_\mathrm{a}(\tau, f_\eta)$ have unit magnitude, which will be a necessary condition in section \ref{sec:quantum-alg} to encode these functions in diagonal unitary matrices.

We show how to compute $H_\mathrm{r}(f_\tau)$ with the POSP \cite{cumming2005digital}. We approximate the integral

\begin{equation}\label{eq:posp-range}
	\begin{split}
		H_\mathrm{r}(f_\tau) = \mathcal{F}[h_\mathrm{r}(\tau)](f_\tau) 
		=
		\int_{- \infty}^{+\infty}
		\chi(\tau) e^{- \mathrm{i} \alpha \pi \tau^2}e^{- 2 \pi \mathrm{i} f_\tau \tau} 
		d\tau
	\end{split}
\end{equation}
by deriving the phase $\theta(\tau, f_\tau) = -\alpha \pi \tau^2 - 2 \pi f_\tau \tau$ with respect to $\tau$ and finding its critical point, which is $\tau_0 = -f_\tau / \alpha $. If we plug the value back into the phase we get $\theta(\tau_0(f_\tau)) = \pi f_\tau^2/\alpha$ and the approximated reference function 

\begin{equation}\label{eq:H_r}
	H_\mathrm{r}(f_\tau) 
	\approx
	\chi(-f_\tau/\alpha) \exp \left\{ \mathrm{i}\pi f_\tau^2/\alpha \right\}
	.
\end{equation}
We can compute $H_\mathrm{m}(f_\tau, f_\eta)$ and $H_\mathrm{a}(\tau, f_\eta)$ with an analogous calculation and obtain 

\begin{align}
	H_\mathrm{m}(f_\tau, f_\eta) \label{eq:rcm-reference-func}
	&\approx 
	\exp \left\{ \mathrm{i} \frac{\pi r_0 c}{2 v^2 f_\tau} f_\eta^2\right\}
	\\
	H_\mathrm{a}(\tau, f_\eta) \label{eq:azimuth-reference-func}
	& \approx
	\exp \left\{
	\mathrm{i}
	\left(
	\frac{4 \pi r_0(\tau) f_0 }{c} 
	-
	\frac{\pi r_0(\tau) c f_\eta^2}{2 v^2 f_0} 
	\right)
	\right\} 
	.
\end{align}
The compressed image is obtained by multiplying the signal by its reference functions. Algorithm \ref{alg:range-Doppler-algorithm} summarizes the classical range-Doppler algorithm.

\begin{algorithm}
	\caption{Classical Range-Doppler Algorithm}\label{alg:range-Doppler-algorithm}
	\begin{algorithmic}
		
		\State $S(f_\tau, f_\eta) = \text{2D-FFT}[s(\tau, \eta)]$
		
		\State $S_\mathrm{r}(f_\tau, f_\eta) = S(f_\tau, f_\eta) \cdot H_\mathrm{r}(f_\tau)$ 
		
		\State $S_{\mathrm{r},\mathrm{m}}(f_\tau, f_\eta) = S_\mathrm{r}(f_\tau, f_\eta) \cdot H_\mathrm{m}(f_\tau, f_\eta)$ 
		
		\State $S_{\mathrm{r},\mathrm{m}, \mathrm{a}}(f_\tau, f_\eta) = S_{\mathrm{r},\mathrm{m}}(f_\tau, f_\eta) \cdot H_\mathrm{a}(\tau, f_\eta)$
		
		\State $s_{\mathrm{r}, \mathrm{m}, \mathrm{a}}(\tau, \eta) = \text{2D-IFFT}[S_{\mathrm{r},\mathrm{m},\mathrm{a}}(f_\tau, f_\eta)]$
		
	\end{algorithmic}
\end{algorithm}

\section{Quantum Range-Doppler Algorithm}\label{sec:quantum-alg}

In this section we propose a quantum algorithm that adapts the classical range-Doppler algorithm to the quantum computing framework. The pixels of the raw data are encoded in a quantum state and the information is manipulated in such a way that the output state will contain the information from the focused image. 

As preliminary notation, we define the image as a grid of $N$ total pixels, the product of sides $d_\mathrm{r}$ and $d_\mathrm{a}$ which represent respectively the number of rows and columns. We indicate with $a_{ij}  \in \mathbb{C}$ the pixel in row $i = 0, \dots, d_\mathrm{r}-1$ and column $j = 0, \dots, d_\mathrm{a}-1$. Since the data can always be padded so that the resulting image is square, we assume in the following that $d_\mathrm{a} = d_\mathrm{r} =: d$, so that $N = d \times d$.  We represent the image as a complex matrix where the first index indicates the row (the r, or range, dimension), and the second indicates the column (the a, or azimuth, dimension). 

\subsection{Quantum State Preparation}\label{sec:encoding}
The raw data is embedded into the qubits through amplitude encoding, which consists of encoding the complex values into the amplitudes of the quantum state. This requires $n = \log_2 N$ qubits. If $N$ is not a power of two it can always be zero-padded.

\begin{figure}[htbp]
	\centering
	\includegraphics[width=0.3\textwidth]{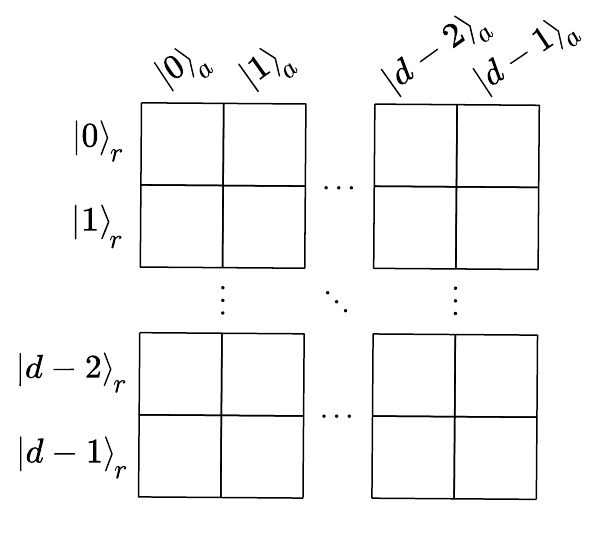}
	\caption{Representation of the amplitude encoding of the complex SAR raw data. $n/2$ qubits encode the range (r) dimension and the remaining $n/2$ qubits encode the azimuth (a) dimension.}
	\label{fig:encoding}
\end{figure}

Before embedding the values in the amplitudes, the complex SAR image is normalized in such a way that the norm of the matrix, thought as a vector of coefficients, is one. This means that every pixel value $a_{ij}$ is normalized according to

\begin{equation}\label{eq:normalization}
	a'_{ij}  = \frac{a_{ij}}{\sqrt{\sum\limits_{h,k = 0}^{d-1} |a_{hk}|^2}}
	\;\;\;\; \forall i,j = 0, \dots, d-1
	.
\end{equation}
From now on, for notational simplicity, we will denote the normalized pixel values with $a_{ij}$. Every complex number $a_{ij}$ is encoded as the amplitude of the state $\ket{i}_\mathrm{r}\ket{j}_\mathrm{a}$, according to Figure~\ref{fig:encoding}. Here the $i$-th row is indexed by the basis element whose binary representation corresponds to the integer $i$, and the same holds for every column $j$.

As it is usually done we adopt the computational basis notation defining
\begin{equation}
	\begin{aligned}
		\ket{0}_\mathrm{r} & := \ket{0,\dots, 0}_\mathrm{r}, \; \dots \; , \ket{d-1}_\mathrm{r} := \ket{1,\dots, 1}_\mathrm{r} \\
		\ket{0}_\mathrm{a} & := \ket{0,\dots, 0}_\mathrm{a}, \; \dots \; , \ket{d-1}_\mathrm{a} := \ket{1,\dots, 1}_\mathrm{a}
		.
	\end{aligned}
\end{equation}
With this notation the state can be rewritten as

\begin{equation}
	\ket{\psi} = 
	\sum\limits_{i,j = 0}^{d-1} a_{ij} \ket{i}_\mathrm{r} \ket{j}_\mathrm{a}, 
\end{equation}
or, in a more informative way, as 

\begin{equation}\label{eq:informative-state}
	\ket{\psi} = \sum\limits_{i}\ket{i}_\mathrm{r}\left(\sum\limits_{j}a_{ij} \ket{j}_\mathrm{a}\right)
	.
\end{equation}
Equation (\ref{eq:informative-state}) can be read as: we select one row $i$, and in that row we have all the values $a_{i, 0}, \dots, a_{i, d-1}$ in the pixels. This notation helps to quickly visualize what information each row or column is containing. 

Finally, we introduce the following notation: with the symbol $\sim$ we refer to the range frequency $(f_{\tau})$ domain, while with $\wedge$ we refer to the azimuth frequency $(f_{\eta})$ domain. This means that the states $\Tilde{\psi}$, $\hat{\psi}$ and $\hat{\Tilde{\psi}}$ indicate that a Fourier transform on $\psi$ has been performed along, respectively, the range, azimuth and both dimensions.

\subsection{Quantum Range Compression}
The first step consists of obtaining the state $\ket{\Tilde{\psi}}$ that contains the Fourier transform of every column, which is

\begin{equation}\label{eq:goal-state-range-migration}
	\ket{\Tilde{\psi}}
	:= 
	\sum_h
	\left( 
	\sum_l \Tilde{a}_{lh} \ket{l}_\mathrm{r}
	\right)
	\ket{h}_\mathrm{a}
	,
\end{equation}
where $\Tilde{a}_{lh} = \sum_i a_{ih} \omega^{li}$ and $\omega = \exp(2 \pi \mathrm{i}/d)$. In equation (\ref{eq:F_r}) we show that the result of equation (\ref{eq:goal-state-range-migration}) can be obtained by applying one single quantum Fourier transform (QFT) on the qubits encoding the r-dimension. Namely

\begin{equation}
	\ket{\Tilde{\psi}} = F_\mathrm{r} \otimes \mathbb{I}_\mathrm{a} \ket{\psi}
	.
\end{equation}
The next step is the multiplication by the range compression filter in frequency domain of equation ($\ref{eq:H_r}$), which yields the state 

\begin{equation}
	\ket{\Tilde{\psi}_\mathrm{r}} 
	:= 
	\sum_h \left( \sum_l \Tilde{\phi}_l \Tilde{a}_{l h} \ket{l}_\mathrm{r} \right) \ket{h}_\mathrm{a}
	,
\end{equation}
with 
\begin{equation}
	\Tilde{\phi}_l = \chi(-f^l_\tau/\alpha) e^{\mathrm{i}\pi (f^l_\tau)^2/\alpha },
\end{equation}
for some discrete frequencies $f^0_\tau, \dots, f^{d-1}_\tau$. This shows that every column $\ket{h}_\mathrm{a}$ has been element-wise multiplied by the reference function in range. It is easy to see that $\ket{\Tilde{\psi}_\mathrm{r}}$ can be obtained from $\ket{\Tilde{\psi}}$ applying the operator

\begin{equation}\label{eq:U_r_r-operator}
	U_{\mathrm{r}} 
	= 
	\sum_{m = 0}^{d-1}
	\Tilde{\phi}_m \ket{m}_\mathrm{r} \bra{m}_\mathrm{r}
	=
	\begin{pmatrix} 
		\Tilde{\phi}_0 & \dots  & 0      \\
		\vdots                 & \ddots & \vdots \\
		0                      & \dots  & \Tilde{\phi}_{d-1}
	\end{pmatrix}
	,
\end{equation}
where the subscript $\mathrm{r}$ reminds us of the fact that the operator encodes the range matched filter.
A direct calculation yields

\begin{equation}
	\begin{aligned}
		U_{\mathrm{r}} \ket{\tilde{\psi}}
		& =
		\left( 
		\sum_m \tilde{\phi}_m \ket{m}_\mathrm{r} \bra{m}_\mathrm{r}
		\right)
		\left(
		\sum_{hl} 
		\tilde{a}_{lh} 
		\ket{l}_\mathrm{r}
		\ket{h}_\mathrm{a}
		\right)
		\\
		& =
		\sum_{h}
		\left(
		\sum_l
		\tilde{\phi}_l
		\tilde{a}_{lh} 
		\ket{l}_\mathrm{r}
		\right)
		\ket{h}_\mathrm{a}
		=
		\ket{\tilde{\psi}_\mathrm{r}}
		.
	\end{aligned}
\end{equation}
Because $U_{\mathrm{r}}$ is diagonal, the condition under which it is unitary requires that every entry $\Tilde{\phi}_l$ has absolute value equal to one, namely $|\Tilde{\phi}_l|^2 = 1 \; \forall l$. The POSP applied in equation (\ref{eq:posp-range}) ensures this, which would otherwise not be true in general.

Finally, the inverse quantum Fourier transform (IQFT) $F_\mathrm{r}^{-1} = F^\dagger_\mathrm{r}$ can be applied to the r-qubits to obtain the time domain state $\ket{\psi_\mathrm{r}}$, which is now focused in range. The calculations are analogous to the ones of equation (\ref{eq:F_r}) and yield

\begin{equation}\label{eq:final-state-range}
	\begin{split}    
		\ket{\psi_\mathrm{r}} := F^\dagger_\mathrm{r} \otimes \mathbb{I}_\mathrm{a} \ket{\tilde{\psi}_\mathrm{r}}
		=
		\sum_{h}
		\left(
		\sum_l
		b_{lh} 
		\ket{l}_\mathrm{r}
		\right)
		\ket{h}_\mathrm{a}
		.
	\end{split}
\end{equation}
Algorithm~\ref{alg:quantum-range-compression} summarizes the main steps.

\begin{algorithm}
	\caption{Quantum Range Compression}\label{alg:quantum-range-compression}
	\begin{algorithmic}
		\State $\ket{\Tilde{\psi}} = F_\mathrm{r} \ket{\psi}$
		\State $\ket{\Tilde{\psi}_\mathrm{r}} = U_{\mathrm{r}} \ket{\tilde{\psi}}$
		\State $\ket{\psi_\mathrm{r}} = F^{\dagger}_\mathrm{r} \ket{\tilde{\psi}_\mathrm{r}}$
	\end{algorithmic}
\end{algorithm}

\subsection{Quantum Range Cell Migration Correction}
The first step to perform the correction of the range cell migration is performing a 2D Fourier transform. This can be achieved by applying a QFT once on the r- and once on the a-qubits, that is $F_\mathrm{r} \otimes F_\mathrm{a}$. A direct computation shown in equation (\ref{eq:2d-qft}) yields 

\begin{equation}
	\ket{\hat{\tilde{\psi}}_\mathrm{r}} 
	=
	\sum_{i,j} 
	\hat{\tilde{c}}_{ij}
	\ket{i}_\mathrm{r}\ket{j}_\mathrm{a}
	= 
	F_\mathrm{r} \otimes F_\mathrm{a}
	\ket{\psi_\mathrm{r}}
	,
\end{equation}
with $\hat{\Tilde{c}}_{ij}$ the 2D Fourier coefficients. Under the assumption that the swath width is not too wide \cite{cumming2005digital}, to every column $j$ identified by a given frequency $f^j_\eta$ corresponds a range cell migration correction filter $H_\mathrm{m}(f^j_\tau, f^i_\eta)$, as shown in equation (\ref{eq:rcm-reference-func}), that must be multiplied element-wise. Once again we write it as an operator $U^i_{\mathrm{m}}$, namely
\begin{equation}
	U^i_{\mathrm{m}} 
	=
	\sum_{j = 0}^{d-1}
	\hat{\Tilde{\gamma}}^i_j
	\ket{j}_\mathrm{r} \bra{j}_\mathrm{r}
	=
	\begin{pmatrix} 
		\hat{\tilde{\gamma}}^i_{0} & \dots  & 0      \\
		\vdots                 & \ddots & \vdots \\
		0                      & \dots  & \hat{\Tilde{\gamma}}^i_{d-1}
	\end{pmatrix}
	,
\end{equation}
with coefficients
\begin{equation}
	\hat{\Tilde{\gamma}}^i_{j} 
	=
	\exp\left\{
	\mathrm{i} 
	\frac{\pi r_0 c}{2 v^2 f^j_\tau} (f^i_\eta)^2
	\right\}.
\end{equation}
Every $U^i_{\mathrm{m}}$ is applied to the corresponding state indexed by $\ket{i}_\mathrm{a}$, which can be done by controlling the gate on the a-qubits. We define the multi-controlled multi-target (MCMT) gate

\begin{equation}\label{eq:CU-i}
	C^i_\mathrm{a} U^{i}_{\mathrm{m}}  
	= 
	U^{i}_{\mathrm{m}} 
	\otimes
	\ket{i}_\mathrm{a} \bra{i}_\mathrm{a} 
	+
	\sum\limits_{k \neq i} 
	\mathbb{I}_\mathrm{r}
	\otimes
	\ket{k}_\mathrm{a} \bra{k}_\mathrm{a}
	,
\end{equation}
which applies the gate $U^{i}_{\mathrm{m}}$ if the state encoded in the a-qubits is $\ket{i}_\mathrm{a}$ and does nothing otherwise. In equation (\ref{eq:MCMT-applied}) we show how applying a series of such gates, namely

\begin{equation}
	\ket{\hat{\tilde{\psi}}_{\mathrm{r}, \mathrm{m}}} 
	=        
	C^{d-1}_\mathrm{a} U^{d-1}_{\mathrm{m}}
	\dots
	C^0_\mathrm{a} U^{0}_{\mathrm{m}}
	\ket{\hat{\Tilde{\psi}}_\mathrm{r}}
\end{equation}
yields a state that has been element-wise multiplied by the appropriate reference function. In Appendix \ref{sec:appendix-mcmt} we also show how the general controlled gate $C^i_\mathrm{a} U^{i}_{\mathrm{m}}$ can be built from the usual multi-controlled gate on the state $\ket{d-1}_\mathrm{a}$ by first swapping the states $\ket{i}_\mathrm{a}$ and $\ket{d-1}_\mathrm{a}$ through a series of $X$-gates, then applying $C^{d-1}_\mathrm{a} U^{i}_{\mathrm{m}}$ and finally applying the same series of $X$-gates to revert the states to its original value.

Finally, a 2D IQFT is applied on the a- and r- dimensions to obtain the signal in time domain as

\begin{equation}
	\ket{\psi_{\mathrm{r}, \mathrm{m}}}
	:=
	F^\dagger_\mathrm{r} \otimes F^\dagger_\mathrm{a} \ket{\hat{\tilde{\psi}}_{\mathrm{r}, \mathrm{m}}} 
	= 
	\sum_{i,j} 
	d_{ij}
	\ket{i}_\mathrm{r}\ket{j}_\mathrm{a}
	.
\end{equation}
The complete procedure is shown in Algorithm \ref{alg:quantum-range-cell-migration-correction}.

\begin{algorithm}[H]
	\caption{Quantum Range Cell Migration Correction}\label{alg:quantum-range-cell-migration-correction}
	\begin{algorithmic}
		\State $\ket{\hat{\tilde{\psi}}_\mathrm{r}} = F_\mathrm{r} \otimes F_\mathrm{a} \ket{\psi_\mathrm{r}} $ 
		\For{every column $\ket{i}_\mathrm{a}
			$ in azimuth}
		\State $\ket{\hat{\Tilde{\psi}}_{\mathrm{r}, \mathrm{m}}} = C^i_\mathrm{a} U^i_{\mathrm{m}}  \ket{\hat{\Tilde{\psi}}_\mathrm{r}}$
		\EndFor
		\State $\ket{\psi_{\mathrm{r}, \mathrm{m}}} = F^\dagger_\mathrm{r} \otimes F^\dagger_\mathrm{a} \ket{\hat{\tilde{\psi}}_{\mathrm{r}, \mathrm{m}}}$
	\end{algorithmic}
\end{algorithm}

\subsection{Quantum Azimuth Compression}

To compress the signal in azimuth we proceed analogously to the range cell migration correction steps, since in this case there is a different reference function for every azimuth row. We first apply an azimuth QFT, thus obtaining

\begin{equation}
	\ket{\hat{\psi}_{\mathrm{r}, \mathrm{m}}}
	=
	\mathbb{I}_\mathrm{r}  \otimes F_\mathrm{a}
	\ket{\psi_{\mathrm{r}, \mathrm{m}}}
	=
	\sum_{i}
	\ket{i}_\mathrm{r} 
	\left(
	\sum_j
	\hat{d}_{ij}
	\ket{j}_\mathrm{a}
	\right)
	.
\end{equation}
The reference function of equation (\ref{eq:azimuth-reference-func}) is encoded in the unitary 

\begin{equation}
	U^i_{\mathrm{a}} 
	=
	\sum_{j = 0}^{d-1}
	\hat{\delta}^i_j
	\ket{j}_\mathrm{a} \bra{j}_\mathrm{a}
	=
	\begin{pmatrix} 
		\hat{\delta}^i_{0} & \dots  & 0      \\
		\vdots             & \ddots & \vdots \\
		0                  & \dots  & \hat{\delta}^i_{d-1}
	\end{pmatrix}
	,
\end{equation}
with coefficients 
\begin{equation}
	\hat{\delta}^i_j = 
	\exp\left\{ \mathrm{i} \left(
	\frac{4 \pi \tau_i f_0 }{2} 
	-
	\frac{\pi \tau_i c^2 (f^j_\eta)^2}{4 v^2 f_0} 
	\right)
	\right\}.
\end{equation}
We construct the multi-controlled multi-target unitary $C^i_\mathrm{r} U^i_{\mathrm{a}}$ which applies $U^i_{\mathrm{a}}$ if the qubits encoding the r-dimension are in the state $\ket{i}_\mathrm{r}$. The desired state can be obtained applying the sequence of MCMT gates

\begin{equation}\label{eq:W-operator-azimuth}
	\ket{\hat{\psi}_{\mathrm{r}, \mathrm{m}, \mathrm{a}}} 
	=        
	C^{d-1}_\mathrm{r} U^{d-1}_{\mathrm{a}}
	\dots
	C^0_\mathrm{r} U^{0}_{\mathrm{a}}
	\ket{\hat{\psi}_{\mathrm{r},\mathrm{m}}}
	.
\end{equation}
Finally, applying the IQFT on the a-qubits yields the focused SAR image. 
The complete algorithm is shown in Algorithm \ref{alg:quantum-azimuth-compression}.

\begin{algorithm}[H]
	\caption{Quantum Azimuth Compression}\label{alg:quantum-azimuth-compression}
	\begin{algorithmic}
		\State $\ket{\hat{\psi}_{\mathrm{r}, \mathrm{m}}} = F_\mathrm{a} \ket{\psi_{\mathrm{r}, \mathrm{m}}} $ 
		\For{every row $\ket{i}_\mathrm{r}$ in range}
		\State $\ket{\hat{\psi}_{\mathrm{r}, \mathrm{m}, \mathrm{a}}} = C^i_\mathrm{r} U^{i}_{\mathrm{a}} \ket{\hat{\psi}_{\mathrm{r}, \mathrm{m}}}$
		\EndFor
		\State $\ket{\psi_{\mathrm{r}, \mathrm{m}, \mathrm{a}}} = F^\dagger_\mathrm{a} \ket{\hat{\psi}_{\mathrm{r}, \mathrm{m}, \mathrm{a}}}$
	\end{algorithmic}
\end{algorithm}
\subsection{Circuit and Measurements}\label{sec:measurements}
\begin{figure*}
	\centering
	\includegraphics[width=500pt]{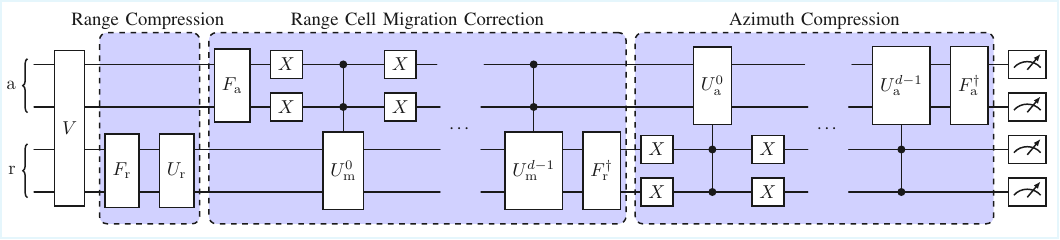}  
	\caption{Complete circuit implementing the proposed quantum range-Doppler algorithm as described in algorithm \ref{alg:complete-quantumrange-Doppler-algorithm}. From the left, it consists of the state preparation gate $V$, and the three blocks of gates contained in the dashed boxes implementing the range compression, the range cell migration correction and the azimuth compression. Finally measurements on every qubit are added to retrieve the probability of every bit string, encoding the value of each pixel of the SAR image. The redundant gates, consisting in adjacent $X$-gates, can be omitted.}
	\label{fig:complete-circuit}
\end{figure*}

Combining the subroutines presented as Algorithms \ref{alg:quantum-range-compression}, \ref{alg:quantum-range-cell-migration-correction} and \ref{alg:quantum-azimuth-compression} we obtain the quantum circuit in Figure \ref{fig:complete-circuit} performing the complete quantum range-Doppler algorithm. Here, the first gate $V$ represents the state preparation circuit, which allows encoding the complex raw data samples into a quantum state. The quantum range compression with gates acting only on the r-qubits follows. Then the quantum range cell migration correction is performed through a sequence of $C^i_\mathrm{a} U^{i}_{\mathrm{m}}$ gates, which have been constructed with a series of $X$-gates and a controlled gate on the state $\ket{i}_\mathrm{a}$, as described in Appendix \ref{sec:appendix-mcmt}. Similarly, the series of $C^i_\mathrm{r} U^{i}_{\mathrm{a}}$ gates are placed for the quantum azimuth compression step. Concatenating Algorithms \ref{alg:quantum-range-compression}, \ref{alg:quantum-range-cell-migration-correction} and \ref{alg:quantum-azimuth-compression} it becomes evident that the QFTs followed by IQFTs are redundant, both on the a- and r-qubits. The same holds for some of the $X$-gates placed between two neighboring multi-controlled gates. These redundancies can be removed by omitting the sequence $F F^\dagger$, in both the a- and r-qubits, and by performing an exclusive or (XOR) operation on the bitstrings of the sequences of $X$-gates. The complete algorithm can be written as shown in Algorithm \ref{alg:complete-quantumrange-Doppler-algorithm}.

\begin{algorithm}
	\caption{Quantum Range-Doppler Algorithm}\label{alg:complete-quantumrange-Doppler-algorithm}
	\begin{algorithmic}
		
		\State $\ket{\tilde{\psi}} = F_\mathrm{r} \otimes \mathbb{I}_\mathrm{a} \ket{\psi} $ 
		
		\State $\ket{\tilde{\psi}_\mathrm{r}} = U_{\mathrm{r}} \otimes \mathbb{I}_\mathrm{a} \ket{\tilde{\psi}}$
		\State $\ket{\hat{\tilde{\psi}}_\mathrm{r}} = \mathbb{I}_\mathrm{r} \otimes F_\mathrm{a} \ket{\tilde{\psi}_\mathrm{r}} $ 
		\For{every column $\ket{i}_\mathrm{a}$ in azimuth}
		\State $\ket{\hat{\tilde{\psi}}_{\mathrm{r}, \mathrm{m}}} = C^i_\mathrm{a} U^i_{\mathrm{m}}  \ket{\hat{\tilde{\psi}}_{\mathrm{r}}}$
		\EndFor
		\State $\ket{\hat{\psi}_{\mathrm{r}, \mathrm{m}}} = F^\dagger_\mathrm{r} \otimes \mathbb{I}_\mathrm{a} \ket{\hat{\tilde{\psi}}_{\mathrm{r}, \mathrm{m}}}$
		\For{every row $\ket{i}_\mathrm{r}$ in range}
		\State $\ket{\hat{\psi}_{\mathrm{r}, \mathrm{m}, \mathrm{a}}} = C^i_\mathrm{r} U^{i}_{\mathrm{a}} \ket{\hat{\psi}_{\mathrm{r}, \mathrm{m}}}$
		\EndFor
		
		\State $\ket{\psi_{\mathrm{r}, \mathrm{m}, \mathrm{a}}} = \mathbb{I}_\mathrm{r} \otimes F^{\dagger}_\mathrm{a} \ket{\hat{\psi}_{\mathrm{r}, \mathrm{m}, \mathrm{a}}} $

	\end{algorithmic}
\end{algorithm}

\newcommand{\multilineustick}[4]{%
	\ustick{%
		\raisebox{#2}{
			\hspace{#1}
			\shortstack{%
				\text{\small #3} \\ 
				\text{\small #4}    
			}%
		}%
	}%
}

Running the circuit multiple times one can extract the probabilities associated to the amplitudes. Suppose that the state $\ket{i}_\mathrm{r}\ket{j}_\mathrm{a}$ has the amplitude $\alpha_{ij}$. By performing multiple measurements we obtain the probability associated to the state, which is $p_{ij} = |\alpha_{ij}|^2$. In order to obtain the modulus of the given pixel we take the square root of the probability $\sqrt{p_{ij}} = | \alpha_{ij} |$. The chosen encoding ensures that the pixels with the highest values, meaning in this case the highest reflectance, are the ones associated to higher probabilities.

\section{Results}\label{sec:results}

\begin{figure}[t]
	\centering
	\includegraphics[width=1\columnwidth]{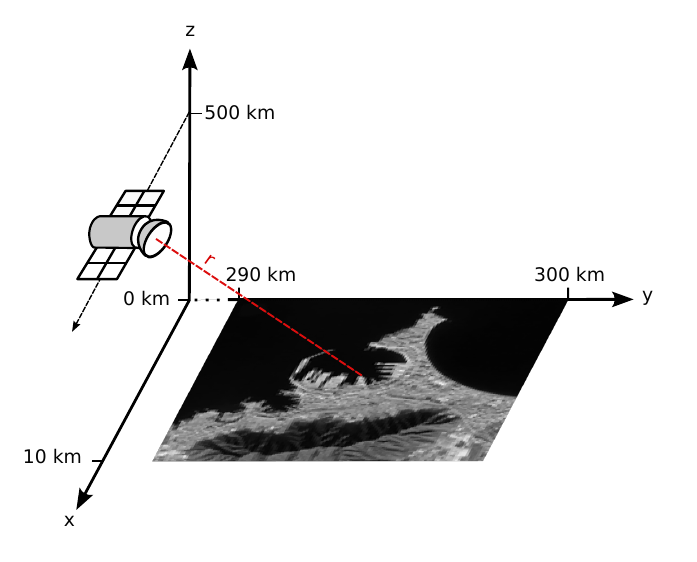}
	\caption{Geometric setup of the simulation. The sensor starts at $\mathbf{x}_0 = (0, 0, \SI{500}{km})$, observing a planar scene at $z=0$ over $x \in [\SI{0}{km}, \SI{10}{km}]$ and $y \in [\SI{290}{km}, \SI{300}{km}]$. The azimuth dimension, which corresponds to the x axis, and range (r) dimensions are depicted. Simulation parameters are summarized in Table \ref{tab:sim-params}.}
	\label{fig:setup}
\end{figure}

\begin{figure}[htbp]
	\begin{minipage}{0.975\columnwidth} 
		\centering
		\includegraphics[width=1\columnwidth]{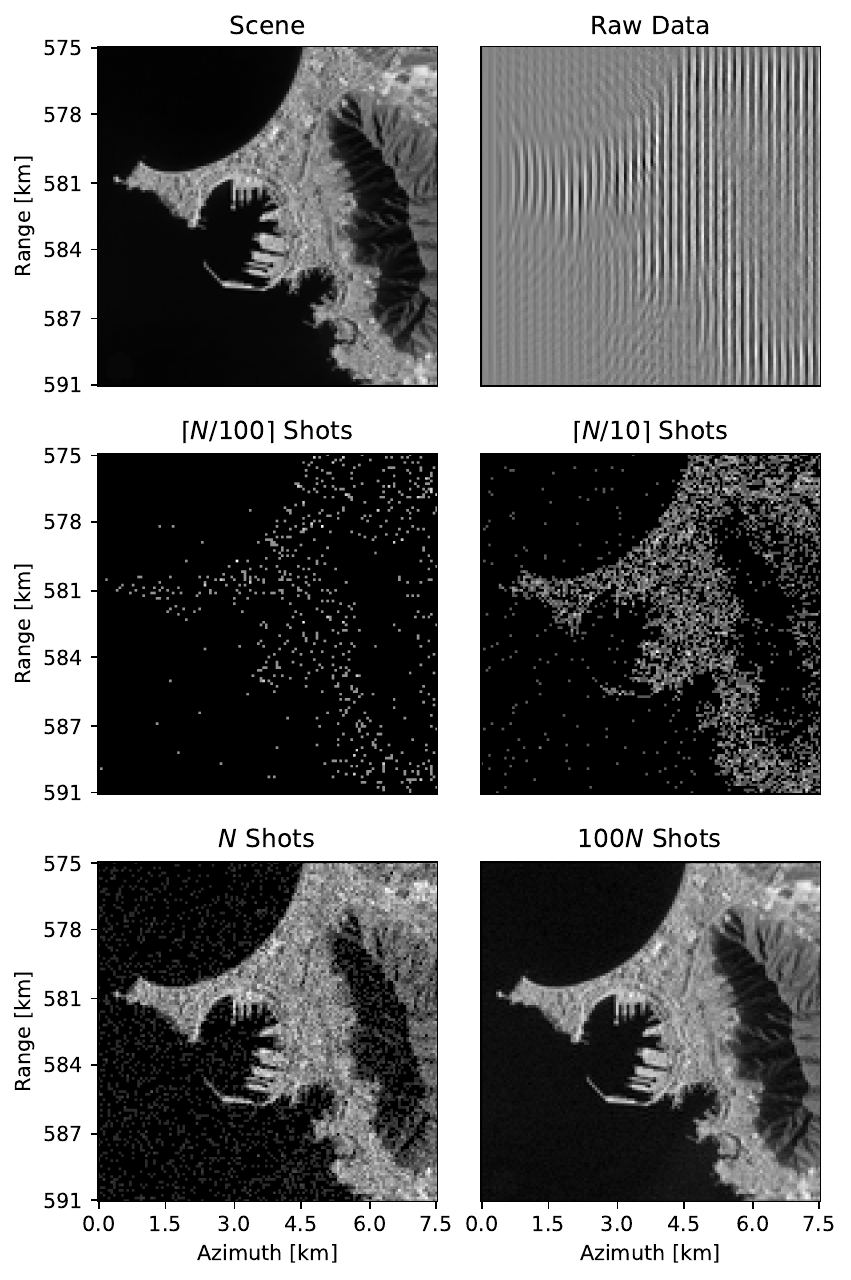}  
		\caption{Results of the simulation for the quantum range-Doppler algorithm for a custom scene taken from \cite{geowgs84WhatImagery} and resized to $N = 128 \times 128$ pixels, consisting of an urban area with a harbor. Upper row: original scene placed on the ground (left) and real part of the complex raw data signals collected by the simulated carrier flying over the scene (right). Central and bottom row: results obtained with the quantum range-Doppler algorithm with $\lceil N/100 \rceil$, $\lceil N/10 \rceil$, $N$  and $100N$ measurements.}
		\label{fig:results}
	\end{minipage}
\end{figure}

The SAR raw data shown in the upper right plot of Figure \ref{fig:results} has been simulated on the basis of the scene presented in the upper left plot. In Figure \ref{fig:setup} the geometric setup of the simulation can be seen: the sensor has been placed in an initial position of $\mathbf{x}_0 = (0, 0, \SI{500}{km})$ with respect to the origin of the axis. The scene has been placed on a surface approximated as a plane at $z=0$ and placed on a grid in the x interval $[\SI{0}{km}, \SI{10}{km}]$ and the y interval $[\SI{290}{km}, \SI{300}{km}]$. We report in Table \ref{tab:sim-params} the parameters of equation (\ref{eq:signal-raw-data}) as well as the other parameters necessary for the raw data simulation. The table shows the velocity $v$ of flight of the carrier along the $x$-axis, which is the azimuth dimension, the flight time $T_\mathrm{f}$, the bandwidth in range $B_\mathrm{r}$, the length $T$ of the pulse, the carrier frequency $f_0$, the reference distance $r_0$, the width $\delta r$ of the scene in range. The amplitude $A_0$ of each point on the ground has been set as the intensity of the corresponding pixel in the scene image.

\begin{table}[htbp]
	\caption{Simulation Parameters}
	\centering
	\setlength{\tabcolsep}{3pt} 
	\scalebox{0.85}{
		\def\arraystretch{1.4} 
		\begin{tabular}{|c|c|c|c|c|c|c|}
			\hline
			$v$ [m/s] & $T_\mathrm{f}$ [s] & $B_\mathrm{r}$ [Hz] & $T$ [s] & $f_0$ [Hz] & $r_0$ [m] & $\delta r$ [m] \\ 
			\hline
			$1.0 \times 10^4$ & 1.0 & $7.4 \times 10^6$ & $8.0 \times 10^{-8}$ & $3.8 \times 10^8$ & $5.8 \times 10^5$ & $5.0 \times 10^3$ \\
			\hline
		\end{tabular}
	}
	\label{tab:sim-params}
\end{table}	
The parameters have been chosen in such a way that the resulting SAR raw data represents a $d \times d$ square image with a total amount of $N = d^2$ pixels that can be encoded in $n = \log_2{N}$ qubits. With the parameters given in Table  \ref{tab:sim-params} the SAR raw data set has $d=128$ samples in range and in azimuth, resulting in $N = 16384$ complex-valued pixels. Some of these parameters are a few orders of magnitude away from realistic scenarios. The TerraSAR-X mission \cite{1025134, 1256959}, for example, uses a signal bandwidth in the order of $B_\mathrm{r} = \SI{1.5e8}{\hertz}$ and a carrier frequency $f_0 =  \SI{9.65e9}{\hertz}$.

The complex raw data samples have been encoded in the input state of the quantum circuit, which has been simulated in Qiskit with the architecture described in Figure \ref{fig:complete-circuit}. In Figure \ref{fig:results} the results of the simulation are shown: in the  top row the plots show on the left the scene placed on the ground used for the simulation that represents an urban area with its harbor, which was taken from \cite{geowgs84WhatImagery} and resized to $N = 128 \times 128$ pixels so to use $\log_2{\left(128^2\right)} = 14$ qubits. On the right the real part of the complex raw data signals is shown. The raw data has been obtained from the SAR sensor flying over the scene with the parameters specified in Table \ref{tab:sim-params}.

The central and bottom rows show the results of the quantum range-Doppler algorithm when $\lceil N/100 \rceil$, $\lceil N/10 \rceil$, $N$ and $100N$ measurements have been performed, where each pixel intensity has been computed as described in Section \ref{sec:measurements}. We can see that, as the number of measurements increases, the results of the quantum range-Doppler algorithm converge to the original scene. 

\section{Complexity}\label{sec:complexity}

We study and compare the theoretical computational complexity of the classical and quantum focusing algorithms in terms of the big-$O$ notation, since we consider the case of an increasing number of pixels in the SAR image. We denote by $N$ the total number of pixels and by $n = \log_2 N$ the number of qubits. The complexities of the classical and quantum algorithms will be studied separately.

\subsection{Classical Algorithm Complexity}
Algorithm \ref{alg:range-Doppler-algorithm} consists of Fourier transforms and element-wise multiplications of the signals by their respective reference functions in the frequency domain. The leading complexity of the algorithm is given by the Fast Fourier Transform (FFT), which has complexity $O(M \log M)$, with $M$ the length of the function it is acting on. In our case $M = \sqrt{N}$, and the FFT is applied to each of the $\sqrt{N}$ columns and rows. The total complexity is then $O(2 \sqrt{N} \cdot\sqrt{N} \log{\sqrt{N}}) = O(N \log{N})$, where the factor two is due to the fact that the FFT is two dimensional, acting on both the rows and the columns. $O(N \log N)$ is then the leading order of the complexity for the classical range-Doppler algorithm.

\subsection{Quantum Algorithm Complexity}
To analyze the complexity of the quantum algorithm, we distinguish between two contributions: the \emph{quantum complexity} \(O_\mathrm{q}\), arising from the decomposition of multi-qubit gates into single- and two-qubit gates, and the \emph{classical complexity} \(O_\mathrm{c}\) that might be necessary as a classical step for the gate decomposition. We examine these complexity contributions separately, focusing on state preparation, algorithmic processing, and measurement.

\subsubsection{State Preparation} 
The state preparation complexity highly depends on its decomposition. 
State preparation algorithms with complexity $O_\mathrm{q}(N)$ that do not make use of ancilla qubits have been proposed \cite{arxiv3, controlledstate}. Adding ancilla qubits \cite{Araujo2021} this can be reduced. Nonetheless this can come at a great cost in terms of resources, since the number of ancilla qubits could grow exponentially with the input size \cite{PhysRevLett.129.230504}. The complexity can also be further reduced \cite{PhysRevA.110.032609} to $O_\mathrm{q}(Sn^2)$ when the initial state is $S$-sparse, meaning that the non-zero entries are at most $S$.

\subsubsection{The Algorithm}
The algorithm itself consists of the three steps previously shown, namely range compression, range cell migration correction and azimuth compression. 

Algorithm \ref{alg:quantum-range-compression} starts and ends, respectively, with a QFT and an IQFT, whose complexities are $O_\mathrm{q}(n^2) = O_\mathrm{q}(\log^2_2 N)$ \cite{Nielsen_Chuang_2010}. An algorithm has been proposed \cite{1306.3991} to decompose diagonal unitary matrices through Walsh operators with optimal gate count $O_\mathrm{q}(2^m)$, where $m$ is the number of qubits they act on. For $U_\mathrm{r}$ we have $m = n/2$, so that $O_\mathrm{q}(2^\frac{n}{2}) = O_\mathrm{q}(\sqrt{N})$. This algorithm requires a classical preprocessing step necessary to compute the $2^m$ Walsh coefficients, which  can be done efficiently in $O(m 2^m) = O(\sqrt{N} \log N)$ through the Fast Walsh–Hadamard transform and yields an additional one-time computation with complexity of order $O_c(\sqrt{N} \log N)$.

Algorithm \ref{alg:quantum-range-cell-migration-correction} applies two QFTs, which contribute with $O(2n^2) = O(\log^2_2 N)$. The sequence of MCMT gates $C^i_\mathrm{a} U^{i}_{\mathrm{m}}$ can be rewritten to obtain a unitary diagonal matrix. We have in fact that
\begin{equation}
	\begin{split}
		C^{d-1}_\mathrm{a} U^{d-1}_{\mathrm{m}} \dots C^0_\mathrm{a} U^{0}_{\mathrm{m}}
		&=
		\sum_{k=0}^{d-1} U^{k}_{\mathrm{m}} \otimes \ket{k}_\mathrm{a}\bra{k}_\mathrm{a}
		.
	\end{split}
\end{equation}
Since every $U^{i}_{\mathrm{m}}$ and every $\ket{k}_\mathrm{a}\bra{k}_\mathrm{a}$ are diagonal, their tensor product is also diagonal, and consequently the whole quantum range cell migration step, performed by the operators $C^{d-1}_\mathrm{a} U^{d-1}_{\mathrm{m}} \dots C^0_\mathrm{a} U^{0}_{\mathrm{m}}$, can be written as a diagonal unitary matrix. We can use the same algorithm presented for the range focusing subroutine and optimally decompose the unitary diagonal matrix with complexity $O_\mathrm{q}(2^m)$. Here $m=n$, thus yielding $O_\mathrm{q}(N)$ and a classical preprocessing complexity of $O_\mathrm{c}(N \log N)$. Although the classical preprocessing introduces a cost of $O_\mathrm{c}(N \log N)$, which is the same as the leading order of the classical range-Doppler algorithm, this is a one-time operation: any subsequent raw datasets using the same antenna geometry and physical setup can be processed without repeating this computation and thus benefiting from the quantum advantage.

Algorithm \ref{alg:quantum-azimuth-compression} exhibits the same structure and operations as Algorithm \ref{alg:quantum-range-cell-migration-correction} but reversed in the control and target qubits. The complexity is thus the same and once again the leading order is $O_\mathrm{q}(N)$ and a classical operation with complexity $O_\mathrm{c}(N \log N)$ which must be carried out once.

In conclusion, the leading order quantum complexity of the proposed algorithm is $O_\mathrm{q}(N)$, which is lower than the complexity of its classical counterpart.

\subsubsection{Measurements}
The proposed algorithm encodes the intensity of each image pixel in the amplitude of the corresponding quantum state. In order to extract this information multiple measurements have to be carried out. Characterizing the probability distribution of $N$ outcomes with statistical significance requires $O(N)$ measurements. Considering that the leading order of the quantum circuit is $O_\mathrm{q}(N)$, this yields a leading order for the total quantum complexity of $O_\mathrm{q}(N^2)$, which is higher than the classical algorithm. 

\section{Discussion and Conclusion}

In this paper we proposed a quantum algorithm that processes SAR raw data to a focused image. In particular, we showed how to map the classical range-Doppler focusing algorithm to the quantum computing framework. In doing this, we discussed how the reference functions, a key ingredient of many focusing algorithms, can be encoded in multi-controlled multi-target unitary operators. 
Unitarity of the compression operations has been achieved by exploiting the principle of stationary phase. We also showed how the proposed encoding, which arises naturally when thinking of the data as a matrix, allows a one- or two-dimensional QFT to act on every row and column in parallel.  

The results of the simulated algorithm show the compression of the raw data signals which encode the information to reconstruct a custom scene. An increasing number of measurements produces a  ``probability map'' which visually outputs the most reflective (most probable) targets before the less reflective (less probable). Both the raw data and the quantum circuit have been simulated, thus ignoring external noise and assuming a fault tolerant framework of computation. 
We found that the computational complexity, considered as the number of single- and two-qubit gates required by the quantum range-Doppler algorithm, is of order $O(N)$. This is true if we exclude the measurements, which are only necessary if the SAR image needs to be visualized but not if it has to be further processed, which is usually the case. This complexity is lower than that of the classical range-Doppler algorithm, namely $O(N \log N)$. When also taking into account classical preprocessing to decompose the unitary multi target gates, given the currently known algorithms available in the literature for gate decomposition, an extra classical complexity $O_\mathrm{c}(N \log N)$ must be taken into account. This contribution to the complexity is as high as the leading order complexity of the range-Doppler algorithm. Nonetheless we stress that the classical preprocessing step must be carried out only once and it depends on the physical parameters of the antenna setup. This means that multiple SAR raw data sets can be processed without repeating this step, taking advantage of the quantum speedup. Thus, the amortized final complexity is $O(N)$.

When dealing with the extra $O_q(N)$ contribution to the complexity due to the measurements one can consider that, due to the encoding method, the most probable pixels correspond to the brightest. As it has been proposed in \cite{Waller2023}, one can then choose a constant number of measurements depending on the amount $M$ of point scatterers that must be visualized. If the desired signal-to-noise ratio is SNR, then we can run the algorithm $\text{SNR} \cdot M$ times, which is a constant factor and thus does not affect the computational complexity. This solution is particularly useful in those scenarios where objects approximated by single point scatterers have to be spotted in an otherwise low reflectivity scene. One example of such a task is ship detection, which is an important field in SAR applications. 

We stress that, in many concrete applications there is no need to visually extract the information since the focused scene contained in the output state will be further processed. One example could be quantum machine learning algorithms where target detection on the encoded inputs is performed \cite{2007.15110}. In these cases the circuit can be carried out only once.

In conclusion, the simulated results confirm that the proposed quantum range-Doppler algorithm effectively compresses and processes SAR raw data while maintaining a computational complexity of order $O_\text{q}(N)$ under ideal conditions. Although classical preprocessing for gate decomposition is as high as the leading order of the classical range-Doppler algorithm, this step is required only once, preserving the algorithm’s amortized complexity. Furthermore, the measurement process allows for a visualization depending on the application requirements, from full-scene reconstruction to selective detection of sparse targets. These properties indicate that the algorithm could serve as a useful step toward the integration of quantum processing techniques into SAR imaging and related remote sensing tasks.

\appendices
\section{Quantum Fourier Transforms}
\renewcommand{\theequation}{A.\arabic{equation}}
\setcounter{equation}{0}

We show that by applying the QFT only once on the qubits encoding the r-dimension we obtain the discrete Fourier transform (DFT) of all the columns of the raw data matrix. We call $F_\mathrm{r}$ the QFT operator acting only on the r-qubits. Then, by the definition of the QFT we have 

\begin{equation}\label{eq:F_r}
	\begin{aligned}
		F_\mathrm{r} \otimes \mathbb{I}_\mathrm{a} \ket{\psi}
		& =
		\left( 
		\sum_{l,k} \omega^{lk} \ket{l}_\mathrm{r} \bra{k}_\mathrm{r}
		\right)
		\left(
		\sum\limits_{i,h} a_{ih} \ket{i}_\mathrm{r} \ket{h}_\mathrm{a} 
		\right)
		\\
		& =
		\sum_{l,i,h} 
		\omega^{li}
		a_{ih}
		\ket{l}_\mathrm{r}
		\ket{h}_\mathrm{a}
		\\
		& =
		\sum_h
		\left( 
		\sum_l \left( 
		\sum_i a_{ih} \omega^{li}
		\right)
		\ket{l}_\mathrm{r}
		\right)
		\ket{h}_\mathrm{a}
		\\
		& =
		\sum_h
		\left( 
		\sum_l \Tilde{a}_{lh} \ket{l}_\mathrm{r}
		\right)
		\ket{h}_\mathrm{a}
		=
		\ket{\Tilde{\psi}}
		,
	\end{aligned}
\end{equation}
where we defined $\Tilde{a}_{lh} = \sum_i a_{ih} \omega^{li}$. Similarly, applying a QFT on the a-qubits and one on the r-qubits we are performing a 2D DFT on all the rows and columns:

\begin{equation}\label{eq:2d-qft}
	\begin{aligned}
		\ket{\hat{\Tilde{\psi}}_\mathrm{r}}
		&=
		F_\mathrm{r} \otimes F_\mathrm{a} \ket{\psi_\mathrm{r}}
		\\
		&=
		\left( 
		\sum_{i,j} \omega^{ij} \ket{i}_\mathrm{r} \bra{j}_\mathrm{r}
		\right)
		\otimes
		\left( 
		\sum_{i',j'} \omega^{i'j'} \ket{i'}_\mathrm{a} \bra{j'}_\mathrm{a}
		\right)
		\\
		&\hspace{12pt}
		\left(
		\sum\limits_{h,l} b_{lh} \ket{l}_\mathrm{r} \ket{h}_\mathrm{a} 
		\right)
		\\
		&= 
		\sum_{i,j,i',j'h,l}
		\omega^{ij} 
		\omega^{i'j'}
		b_{lh}
		\delta_{jl}
		\delta_{j'h}
		\ket{i}_\mathrm{r}\ket{i'}_\mathrm{a}
		\\
		&= 
		\sum_{i,i',j,j'}
		\omega^{ij} 
		\omega^{i'j'}
		b_{j j'}
		\ket{i}_\mathrm{r}\ket{i'}_\mathrm{a}
		.
	\end{aligned}
\end{equation}
If we define the coefficients $\hat{\Tilde{c}}_{i,i'} = \sum_{j,j'} \omega^{ij} \omega^{i'j'} b_{j j'}$ and rewrite $i'$ as $j$ we arrive at the state 

\begin{equation}
	\ket{\hat{\Tilde{\psi}}_\mathrm{r}} 
	=
	\sum_{i,j} 
	\hat{\Tilde{c}}_{ij}
	\ket{i}_\mathrm{r}\ket{j}_\mathrm{a}
	.
\end{equation}

\section{Multi-Controlled Multi-Target Gate}\label{sec:appendix-mcmt}
\renewcommand{\theequation}{B.\arabic{equation}}
\setcounter{equation}{0}

We first show how the MCMT gate $C^i_\mathrm{a}U^{i}_{\mathrm{m}}$ of equation (\ref{eq:CU-i}) can be constructed by swapping the state $\ket{i}_\mathrm{a}$ with $\ket{d-1}_\mathrm{a}$ with a sequence of $X$ gates, then applying the MCMT gate $C^{d-1}_\mathrm{a}U^{i}_{\mathrm{m}}$, namely

\begin{equation}\label{eq:CU-d-1}
	C^{d-1}_\mathrm{a} U^{i}_{\mathrm{m}}  
	= 
	U^{i}_{\mathrm{m}} 
	\otimes
	\ket{d-1}_\mathrm{a} \bra{d-1}_\mathrm{a} 
	+
	\sum\limits_{k \neq d-1} 
	\mathbb{I}_\mathrm{r}
	\otimes
	\ket{k}_\mathrm{a} \bra{k}_\mathrm{a} 
	,
\end{equation}
and finally by swapping the states back to the original form with the same sequence of $X$ gates. To swap $\ket{i}_\mathrm{a}$ with $\ket{d-1}_\mathrm{a}$ we apply an $X$ gate on the qubits with a $\ket{0}_\mathrm{a}$ state to turn it into a $\ket{1}_\mathrm{a}$. More precisely, given column $\ket{i}_\mathrm{a} = \ket{i_0, \dots, i_{d-1}}_\mathrm{a}$ we apply to $\ket{i}_\mathrm{a}$ the sequence of gates

\begin{equation}\label{eq:operator-s}
	S^i_\mathrm{a} := X^{1-i_0} \dots X^{1-i_{d-1}}   
	,
\end{equation}
with 
\begin{equation}
	X^{1-i_k} = 
	\begin{cases}
		X, & \text{if $i_k = 0$} \\
		\mathbb{I}, & \text{if $i_k = 1$}
	\end{cases}
	.
\end{equation}
$S^i_\mathrm{a}$ swaps the states as desired, since $S^i_\mathrm{a} \ket{i}_\mathrm{a} = \ket{d-1}_\mathrm{a}$ and $S^i_\mathrm{a} \ket{d-1}_\mathrm{a} = \ket{i}_\mathrm{a}$. We then have 

\begin{equation}
	C^i_\mathrm{a}U^{i}_{\mathrm{m}} = S^i_\mathrm{a} \cdot C^{d-1}_\mathrm{a}U^{i}_{\mathrm{m}} \cdot S^i_\mathrm{a}
	.
\end{equation}

We now show how applying the MCMT gate $C^i_\mathrm{a}U^{i}_{\mathrm{m}}$ of equation (\ref{eq:CU-i}) element-wise multiplies each column $\ket{i}_\mathrm{a}$ by the appropriate reference function encoded in the coefficients $\hat{\Tilde{\gamma}}^i_j$. A direct calculation yields

\begin{equation}\label{eq:MCMT-applied}
	\begin{aligned}
		& 
		C^i_\mathrm{a}U^{i}_{\mathrm{m}} \ket{\hat{\Tilde{\psi}}_\mathrm{r}}
		=
		C^i_\mathrm{a}U^{i}_{\mathrm{m}}
		\left(
		\sum_{h} \left( \sum_l \hat{\Tilde{c}}_{lh}  \ket{l}_\mathrm{r} \right) \ket{h}_\mathrm{a} 
		\right)
		\\
		& =
		C^i_\mathrm{a}U^{i}_{\mathrm{m}}
		\left[
		\left( \sum_l\hat{\Tilde{c}}_{li}  \ket{l}_\mathrm{r} \right) \ket{i}_\mathrm{a}
		+
		\sum_{h \neq i} \left( \sum_l \hat{\Tilde{c}}_{lh}  \ket{l}_\mathrm{r} \right) \ket{h}_\mathrm{a}
		\right]
		\\
		& =
		\left( \sum_l \hat{\Tilde{\gamma}}^i_l \hat{\Tilde{c}}_{li} \ket{l}_\mathrm{r} \right)
		\ket{i}_\mathrm{a}
		+
		\sum_{h \neq i} \left( \sum_l \hat{\Tilde{c}}_{lh} \ket{l}_\mathrm{r} \right)
		\ket{h}_\mathrm{a}
		.
	\end{aligned}
\end{equation}
By applying $C^i_\mathrm{a}U^{i}_{\mathrm{m}} \ket{\hat{\Tilde{\psi}}_\mathrm{r}}$ for every $i = 0, \dots, d-1$ we see that every column gets element-wise multiplied by the appropriate reference function, thus performing the RCM correction. Analogous calculations, where instead of the a- we control on the r-qubits, perform the azimuth compression as described in equation (\ref{eq:W-operator-azimuth}).

\bibliography{main}

\begin{thebibliography}{10}
\providecommand{\url}[1]{#1}
\csname url@samestyle\endcsname
\providecommand{\newblock}{\relax}
\providecommand{\bibinfo}[2]{#2}
\providecommand{\BIBentrySTDinterwordspacing}{\spaceskip=0pt\relax}
\providecommand{\BIBentryALTinterwordstretchfactor}{4}
\providecommand{\BIBentryALTinterwordspacing}{\spaceskip=\fontdimen2\font plus
\BIBentryALTinterwordstretchfactor\fontdimen3\font minus
  \fontdimen4\font\relax}
\providecommand{\BIBforeignlanguage}[2]{{%
\expandafter\ifx\csname l@#1\endcsname\relax
\typeout{** WARNING: IEEEtran.bst: No hyphenation pattern has been}%
\typeout{** loaded for the language `#1'. Using the pattern for}%
\typeout{** the default language instead.}%
\else
\language=\csname l@#1\endcsname
\fi
#2}}
\providecommand{\BIBdecl}{\relax}
\BIBdecl

\bibitem{6504845}
A.~Moreira, P.~Prats-Iraola, M.~Younis, G.~Krieger, I.~Hajnsek, and K.~P.
  Papathanassiou, ``A tutorial on synthetic aperture radar,'' \emph{IEEE
  Geoscience and Remote Sensing Magazine}, vol.~1, no.~1, pp. 6--43, 2013.

\bibitem{cumming2005digital}
I.~Cumming and F.~Wong, \emph{Digital Processing of Synthetic Aperture Radar
  Data: Algorithms and Implementation}, ser. Artech House remote sensing
  library.\hskip 1em plus 0.5em minus 0.4em\relax Artech House, 2005, no. v. 1.

\bibitem{1435823}
\BIBentryALTinterwordspacing
ESA, ``Terrasar-x esa archive,'' 2014. [Online]. Available:
  \url{https://earth.esa.int/eogateway/catalog/terrasar-x-esa-archive}
\BIBentrySTDinterwordspacing

\bibitem{TanDEM}
\BIBentryALTinterwordspacing
DLR, ``Tdx (tandem-x),'' 2010. [Online]. Available:
  \url{https://www.eoportal.org/satellite-missions/tandem-x#summary}
\BIBentrySTDinterwordspacing

\bibitem{ALOS}
\BIBentryALTinterwordspacing
J.~A.~E. Agency, ``Alos-2 project/palsar-2,'' 2014. [Online]. Available:
  \url{https://www.eorc.jaxa.jp/ALOS-2/en/about/palsar2.htm}
\BIBentrySTDinterwordspacing

\bibitem{NISAR}
\BIBentryALTinterwordspacing
I.~NASA, ``Nisar.'' [Online]. Available:
  \url{https://www.eoportal.org/satellite-missions/nisar#nisar-nasa-isro-synthetic-aperture-radar-mission}
\BIBentrySTDinterwordspacing

\bibitem{9944234}
\BIBentryALTinterwordspacing
ESA. [Online]. Available:
  \url{https://www.eoportal.org/satellite-missions/rose-l#rose-l-sar-performance-prediction}
\BIBentrySTDinterwordspacing

\bibitem{SENTINEL}
\BIBentryALTinterwordspacing
------. [Online]. Available:
  \url{https://sentinel.esa.int/documents/247904/349449/S1_SP-1322_1.pdf}
\BIBentrySTDinterwordspacing

\bibitem{rs14051258}
H.~Cruz, M.~Véstias, J.~Monteiro, H.~Neto, and R.~P. Duarte, ``A review of
  synthetic-aperture radar image formation algorithms and implementations: A
  computational perspective,'' \emph{Remote Sensing}, vol.~14, no.~5, 2022.

\bibitem{6471225}
K.-H. Liu, A.~Wiesel, and D.~C. Munson, ``Synthetic aperture radar autofocus
  via semidefinite relaxation,'' \emph{IEEE Transactions on Image Processing},
  vol.~22, no.~6, pp. 2317--2326, 2013.

\bibitem{9173392}
J.~Gao, Y.~Ye, S.~Li, Y.~Qin, X.~Gao, and X.~Li, ``Fast super-resolution 3d sar
  imaging using an unfolded deep network,'' in \emph{2019 IEEE International
  Conference on Signal, Information and Data Processing (ICSIDP)}, 2019, pp.
  1--5.

\bibitem{5414307}
V.~M. Patel, G.~R. Easley, D.~M. Healy, and R.~Chellappa, ``Compressed sensing
  for synthetic aperture radar imaging,'' in \emph{2009 16th IEEE International
  Conference on Image Processing (ICIP)}, 2009, pp. 2141--2144.

\bibitem{NorouziLarki2023}
\BIBentryALTinterwordspacing
S.~Norouzi~Larki, M.~Mosleh, and M.~Kheyrandish, ``Quantum audio steganalysis
  based on quantum fourier transform and deutsch--jozsa algorithm,''
  \emph{Circuits, Systems, and Signal Processing}, vol.~42, no.~4, pp.
  2235--2258, 4 2023. [Online]. Available:
  \url{https://doi.org/10.1007/s00034-022-02208-y}
\BIBentrySTDinterwordspacing

\bibitem{arxiv1}
T.~de~Lima~Silva, L.~Borges, and L.~Aolita, ``Fourier-based quantum signal
  processing,'' 2022.

\bibitem{arxiv2}
Y.~Liu, J.~M. Martyn, J.~Sinanan-Singh, K.~C. Smith, S.~M. Girvin, and I.~L.
  Chuang, ``Toward mixed analog-digital quantum signal processing: Quantum
  ad/da conversion and the fourier transform,'' 2024.

\bibitem{Nielsen_Chuang_2010}
M.~A. Nielsen and I.~L. Chuang, \emph{Quantum Computation and Quantum
  Information: 10th Anniversary Edition}.\hskip 1em plus 0.5em minus
  0.4em\relax Cambridge University Press, 2010.

\bibitem{Waller2023}
\BIBentryALTinterwordspacing
E.~H. Waller, A.~Keil, and F.~Friederich, ``Quantum range-migration-algorithm
  for synthetic aperture radar applications,'' \emph{Scientific Reports},
  vol.~13, no.~1, p. 11436, 7 2023. [Online]. Available:
  \url{https://doi.org/10.1038/s41598-023-38611-x}
\BIBentrySTDinterwordspacing

\bibitem{2504.01832}
K.~A. Salahat, M.~E. Moussawi, and A.~J. Ghandour, ``Quantum meets sar: A novel
  range-doppler algorithm for next-gen earth observation,'' 2025.

\bibitem{POSP}
A.~Zhu, ``An introduction to the theory of oscillatory integrals.''

\bibitem{1025134}
G.~Krieger, M.~Wendler, H.~Fiedler, J.~Mittermayer, and A.~Moreira,
  ``Performance analysis for bistatic interferometric sar configurations,'' in
  \emph{IEEE International Geoscience and Remote Sensing Symposium}, vol.~1,
  2002, pp. 650--652 vol.1.

\bibitem{1256959}
M.~Stangl, R.~Werninghaus, and R.~Zahn, ``The terrasar-x active phased array
  antenna,'' in \emph{IEEE International Symposium on Phased Array Systems and
  Technology, 2003.}, 2003, pp. 70--75.

\bibitem{geowgs84WhatImagery}
A.~Shrivastava, ``What is sar imagery? a complete guide to synthetic aperture
  radar,'' \url{https://www.geowgs84.com}.

\bibitem{arxiv3}
M.~Mottonen, J.~J. Vartiainen, V.~Bergholm, and M.~M. Salomaa, ``Transformation
  of quantum states using uniformly controlled rotations,'' 2004.

\bibitem{controlledstate}
P.~Yuan and S.~Zhang, ``Optimal (controlled) quantum state preparation and
  improved unitary synthesis by quantum circuits with any number of ancillary
  qubits,'' \emph{Quantum Journal}, vol.~7, 3 2023.

\bibitem{Araujo2021}
\BIBentryALTinterwordspacing
I.~F. Araujo, D.~K. Park, F.~Petruccione, and A.~J. da~Silva, ``A
  divide-and-conquer algorithm for quantum state preparation,''
  \emph{Scientific Reports}, vol.~11, no.~1, p. 6329, 3 2021. [Online].
  Available: \url{https://doi.org/10.1038/s41598-021-85474-1}
\BIBentrySTDinterwordspacing

\bibitem{PhysRevLett.129.230504}
X.-M. Zhang, T.~Li, and X.~Yuan, ``Quantum state preparation with optimal
  circuit depth: Implementations and applications,'' \emph{Phys. Rev. Lett.},
  vol. 129, p. 230504, 11 2022.

\bibitem{PhysRevA.110.032609}
\BIBentryALTinterwordspacing
D.~Ramacciotti, A.~I. Lefterovici, and A.~F. Rotundo, ``Simple quantum
  algorithm to efficiently prepare sparse states,'' \emph{Phys. Rev. A}, vol.
  110, p. 032609, 9 2024. [Online]. Available:
  \url{https://link.aps.org/doi/10.1103/PhysRevA.110.032609}
\BIBentrySTDinterwordspacing

\bibitem{1306.3991}
J.~Welch, D.~Greenbaum, S.~Mostame, and A.~Aspuru-Guzik, ``Efficient quantum
  circuits for diagonal unitaries without ancillas,'' 2013.

\bibitem{2007.15110}
P.~B. Weichman, ``Quantum-enhanced algorithms for classical target detection in
  complex environments,'' 2020.

\end{thebibliography}


\end{document}